\documentclass[12pt]{article}
\usepackage[utf8]{inputenc}
\usepackage{graphicx} 
\usepackage{amsmath} 
\usepackage{fancyhdr} 
\usepackage{geometry} 
\usepackage{booktabs}
\usepackage{colortbl}

\usepackage[style=authoryear-ibid,backend=biber,maxbibnames=99,maxcitenames=2,uniquelist=false,isbn=false,url=true,eprint=false,doi=true,giveninits=true,uniquename=init]{biblatex} 
\AtEveryBibitem{
    \clearfield{urlyear}
    \clearfield{urlmonth}
} 
\AtEveryBibitem{\clearfield{month}} 
\AtEveryCitekey{\clearfield{month}} 
\renewbibmacro{in:}{} 

\renewbibmacro*{editorstrg}{
  \printtext[editortype]{%
    \iffieldundef{editortype}
      {\ifboolexpr{
         test {\ifnumgreater{\value{editor}}{1}}
         or
         test {\ifandothers{editor}}
       }
         {\bibcpstring{editors}}
         {\bibcpstring{editor}}}
      {\ifbibxstring{\thefield{editortype}}
         {\ifboolexpr{
            test {\ifnumgreater{\value{editor}}{1}}
            or
            test {\ifandothers{editor}}
          }
            {\bibcpstring{\thefield{editortype}s}}
            {\bibcpstring{\thefield{editortype}}}}
         {\thefield{editortype}}}}}

\renewbibmacro*{byeditor+others}{
  \ifnameundef{editor}
    {}
    {\printnames[byeditor]{editor}%
     \addspace
     \mkbibparens{\usebibmacro{editorstrg}}
     \clearname{editor}%
     \newunit}%
  \usebibmacro{byeditorx}%
  \usebibmacro{bytranslator+others}}
\AtEveryBibitem{%
  \clearlist{language}%
} 
\citetrackerfalse 
\DeclareNameAlias{sortname}{family-given} 
\usepackage[format=plain,
            font=it]{caption} 
\usepackage[english]{babel}
\usepackage{csquotes}

\newcommand\titleofdoc{Facilitating Cooperation in Human-Agent Hybrid Populations through Autonomous Agents} 

\begin{document}

\begin{titlepage}
   \begin{center}
        \vspace*{4cm} 

        \Huge{\titleofdoc} 


        \vspace{0.75cm}
        \large{Hao Guo$^1$, Chen Shen$^2$, Shuyue Hu$^3$, Junliang Xing$^{4, \dag}$, Pin Tao$^4$, \\ Yuanchun Shi$^4$, Zhen Wang$^{1,\dag}$}

        \small{\dag Corresponding author:  zhenwang0@gmail.com, jlxing@tsinghua.edu.cn}
        
        \vspace{1 cm}
        \small{1. School of Mechanical Engineering, and School of Artificial Intelligence, Optics and Electronics (iOPEN), Northwestern Polytechnical University, Xi'an 710072, China\\
        2. Faculty of Engineering Sciences, Kyushu University, Kasuga-koen, Kasuga-shi, Fukuoka 816-8580, Japan\\
        3. Shanghai Artificial Intelligence Laboratory, Shanghai, China\\
        4. Department of Computer Science and Technology, Tsinghua University, Beijing 100084, China}
    
\date{\today}

        \vspace{3 cm}
        \Large{\today}
        

       \vfill
    \end{center}
\end{titlepage}

	


\section*{SUMMARY}
\label{summary-abstract}

Cooperation is a vital social behavior that plays a crucial role in human prosperity, enabling conflict resolution and averting disastrous outcomes. With the increasing presence of autonomous agents (AAs), human-agent interaction becomes more frequent in modern society. We investigate the impact of cooperative and defective AAs on human cooperation within the framework of evolutionary game theory, particularly in one-shot social dilemma games. Our findings reveal that cooperative AAs have a limited impact on prisoner's dilemma, but facilitate cooperation in stag hunt games. Surprisingly, defective AAs, rather than cooperative AAs, promote complete dominance of cooperation in snowdrift games. Meanwhile, in scenarios with weak imitation strength, cooperative AAs are able to maintain or even promote cooperation in all these games. Additionally, the results obtained from structured populations also imply that the effectiveness of AAs in promoting cooperation can be maximized by carefully considering their design and application in a given context.


\section{INTRODUCTION}
\label{intro}

Cooperation, which serves as a fundamental social behavior (\cite{nowak2006five,west_jeb07}), plays a crucial role in ensuring human prosperity. It not only facilitates the resolution of individual conflicts, such as hunting and driving, but also mitigates burdensome catastrophes like global climate change and disease transmission (\cite{vasconcelos2013bottom,bauch2004va}). However, cooperation often struggles to survive in the face of competition with defection due to lower payoffs (\cite{hauert2005game}). Although mutual cooperation is beneficial to collective interests, individuals are frequently tempted to choose defection. The concept of social dilemma captures the inherent challenge in the evolution of cooperation, referring to a situation where an individual's interests conflict with collective interests. Two-player social dilemma games such as prisoner's dilemma (PD) game, stag hunt (SH) game, and snowdrift (SD) game, are employed ubiquitously, portraying the rational decision-making of two participants using a strategy set and payoff matrix  (\cite{perc2017statistical, wang2015universal}). This type of matrix game allows for equilibrium analysis and has been extensively utilized in research within the fields of social science, biology, and artificial intelligence (AI) (\cite{chu2020evolution, szolnoki2015conformity, hu2018social}).

With the integration of AI into various aspects of human life, advancements in science and technology have allowed humans to delegate decision-making tasks to machines (\cite{de2019human, bonnefon2016social, faisal2019understanding}). 
Although previous studies have suggested fascinating solutions to encourage cooperation in human-human interactions (\cite{guo2020novel, chen2015first}), they do not address this problem within human-agent hybrid populations. 
Consequently, research on human-agent coordination has gained significant attention and encompasses diverse areas. One typical example is autonomous driving (\cite{nair2021sharing}), where humans relinquish decision-making power to cars, thereby freeing themselves from the physical demands of driving and making travel easier and more enjoyable. However, most of these researches focus on situations where humans and agents share a \textit{common goal} (\cite{nikolaidis2015efficient, beans2018can}). When conflicts of interest arise, it becomes crucial to investigate the evolution of human behavior in a human-agent hybrid environment (\cite{crandall2018cooperating}). 
As social interactions have become more hybrid (\cite{silver2016mastering,nikolaidis2015efficient}), involving humans and AAs, there lies an opportunity to gain new insights into how human cooperation is affected (\cite{correia2019record,crandall2018cooperating}).
This work aims to examine the influence of AAs on human cooperative behavior when social dilemmas exist.

Understanding how human behavior changes in the presence of robots or AAs is a challenging but essential topic (\cite{sheridan2016human,paiva2018engineering}).
To accurately capture human cooperation toward agents, previous studies have primarily focused on developing (or designing) algorithms for AAs (\cite{crandall2015robust,crandall2018cooperating}). In particular, they mainly focused on repeated games where human players (HPs) can make decisions based on historical information about AAs. The impact of one-shot settings, where players lack prior experience and information about their counterparts, has been generally ignored with few exceptions (\cite{terrucha2022art,santos2019evolution}). In this paper, we focus on how cooperative and defective AAs affect human cooperation in two-player social dilemma games, and we ask: Are cooperative AAs always beneficial to human cooperation? Do defective AAs consistently impede the evolution of human cooperation? How do population dynamics change in structured and unstructured populations when the ratio of human-human interaction to human-agent interaction varies?



\begin{figure}
\centering
\includegraphics[scale=0.4]{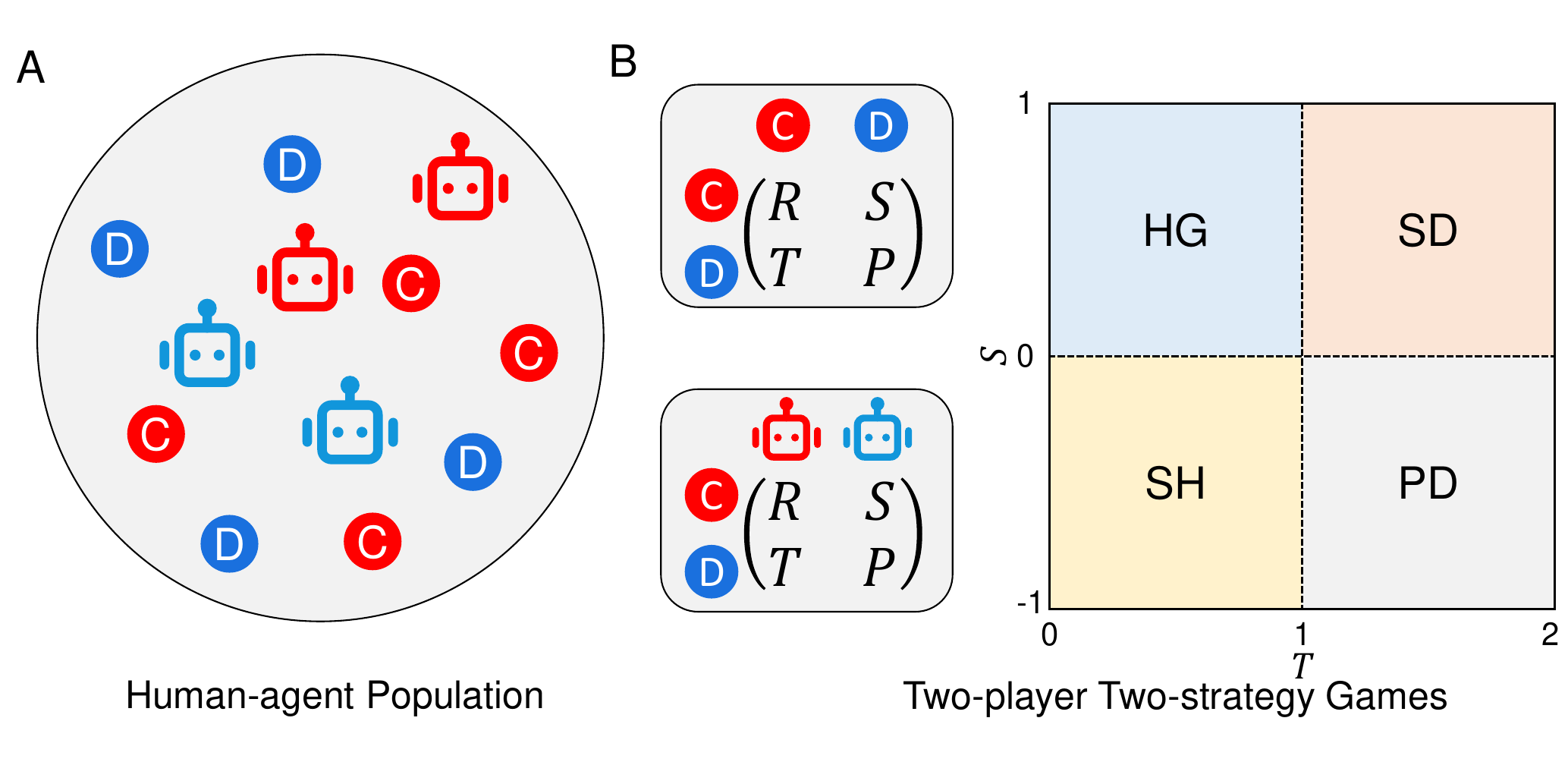}
\caption{ \textbf{Schematic representation of the human-agent hybrid population.} (A) The well-mixed population consists of human players and AAs interacting with each other. The frequencies of human-human and human-agent interactions depend on the composition of the population, namely the ratio of AAs to human players. The red and blue solid circles respectively represent cooperators and defectors in the human players. (B) The two-player two-strategy games in human-human and human-agent interactions. By setting $R=1$ and $P=0$, there are four kinds of games, including prisoner's dilemma game, snowdrift game, stag hunt game, and harmony game.}
\label{sketch}
\end{figure}

To address the research questions mentioned above, we utilize an evolutionary game theoretic framework to study the conundrum of cooperation in social dilemma games with a one-shot setting. As shown in Fig.~\ref{sketch}, the typical games involve PD, SD, and SH games. As previous evidence has proved (or hypothesized), human players update strategies according to payoff differences, with social learning being the most well-known modality (\cite{sigmund2010social, santos2006cooperation, traulsen2007pairwise}). 
Therefore, we examine the evolutionary dynamics of human cooperation by employing replicator dynamics and pairwise comparison (\cite{roca2009evolutionary,traulsen2010human, traulsen2007pairwise}).
The main difference between these two dynamics lies in the consideration of AAs. In the pairwise comparison rule, human players can imitate the strategies of AAs, whereas replicator dynamics do not incorporate such imitation.

In the human-agent hybrid population, the fraction of cooperation among human players is denoted as $\rho_C$ ($0 \leq \rho_C \leq 1$). Assuming human players are one unit, add $y$ units of AAs to the hybrid population. Specifically, AAs are programmed to choose cooperation with a fixed probability $\phi$ that remains constant over time. The summary of the notations is given in Table~\ref{t01}. Our findings indicate that in well-mixed populations with replicator dynamics, AAs have little impact on equilibrium in games with a dominant strategy (Theorem 1 in  electronic supplementary material). Cooperative agents facilitate cooperation in SH games (Theorem 3 in electronic supplementary material) but undermine cooperation in SD games. Counterintuitively, seemingly harmful defective agents can support the dominance of cooperation in SD games (Theorem 2 in  electronic supplementary material). Additionally, we conduct stability analysis and establish the conditions for the prevalence of cooperation. Our results demonstrate that even a minority of defective (or cooperative) AAs can significantly enhance human cooperation in SD (SH) games. However, when cooperative (or defective) AAs constitute a majority, they may trigger the collapse of cooperation in SD (or SH) games. These findings are further corroborated by pairwise comparison rule when strong imitation strength is considered. In scenarios with weak imitation strength, cooperative AAs are more likely to stimulate human cooperation.

In contrast to well-mixed populations, where players can interact with others with an equal probability, structured populations restrict interactions to locally connected neighbors. Such a difference in interactive environments is deemed a determinate factor influencing the emergence of cooperation (\cite{ohtsuki2006simple}).
To investigate this, we conduct experimental simulations on complex networks and observe that structured populations yield comparable results to well-mixed scenarios, except in the case of heterogeneous networks. This divergence can be attributed to the influential role of nodes with higher degrees. Our results, taken together, provide valuable insights into the impact of population state control on humans.


    \begin{table*}
	\centering
	\setlength{\tabcolsep}{6mm}
	\renewcommand \arraystretch{1.5}
	\caption{\label{t01} Summary of notations.}
	\begin{tabular}{cc}
		\toprule
		$y$ & The ratio of autonomous agents to human players   \\
		$\phi$ & The cooperation probability of autonomous agents    \\
		$\rho_C$ & The frequency of human cooperation among human players      \\
		$K$	&	The imitation strength of human player   \\
		\bottomrule
	\end{tabular}
\end{table*}

\section{RESULTS}
\label{tresults}

In this section, we mainly present the theoretic results of well-mixed populations in PD, SD, and SH games by analyzing the replicator dynamics and pairwise comparison rule. 
The replicator equation (\cite{roca2009evolutionary}) is a differential equation, depicting the growth of a specific strategy based on the payoff difference. The pairwise comparison rule depicts the process of strategy imitation according to Fermi function. At last, we present an extension to complex networks.

\subsection{Replicator dynamics}


\subsubsection{Prisoner's dilemma game and harmony game}

\begin{figure}
\centering
\includegraphics[scale=8]{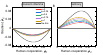}
\caption{\textbf{Phase portrait of PD and HG games.} (A) In PD game, additional AAs affect the convergence rate without altering the equilibrium. (B) In SH game, AAs do not influence the equilibrium but the convergence rate. The baseline means the situation with the absence of AAs. Parameters are fixed as $S=-0.2$, $T=1.3$ for PD game, $T=0.5$, $S=0.1$ for SH game, and $y=0.95$.}
\label{PD}
\end{figure}

In PD game, even though the presence of AAs, defection is the dominant strategy and the expected payoff of defection is equal to or larger than cooperation. Therefore, evolutionary dynamics reach a full defection equilibrium state, and human cooperation diminishes irrespective of its initial frequency (see Theorem 1 in electronic supplementary material). This finding remains robust against any cooperation probability of AAs, as shown in Fig.~\ref{PD} A. Although the equilibrium remains constant, the convergence rate is influenced by the values of $y$ and $\phi$. 
In contrast, the expected payoff of cooperation is equal to or larger than defection (see Fig.~\ref{PD} B) in a harmony game, thereby the system emerges into a full cooperation equilibrium state. 
The results show that both cooperative and defective AAs have no effect on the convergence state if the game has a dominant strategy.

\subsubsection{Snowdrift game}

\begin{figure*}
\centering
\includegraphics[scale=8]{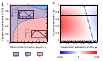}
\caption{\textbf{Equilibrium and phase diagram in SD game.} (A) A phase diagram of stable equilibrium state as a function of $(\phi, y)$ pair. The blue, purple, and red areas mean the full C state, the coexistence state of C and D, and the full D state, respectively. The insets are phase portraits of human cooperation $\rho_C$. In the full C state, the gradient of $\rho_C$ is always not less than 0. In the region where cooperation and defection coexist, the gradient of $\rho_C$ is non-negative when $\rho_C<\rho_C^*$, and non-positive otherwise. The dashed line $y=1$ indicates an equal contribution of human players and AAs, each accounting for 0.5. (B) The stable (solid dots) and unstable (open dots) equilibrium states as a function of $\phi$ with $y=4$. The background means the gradient of $\rho_C$, which reveals the evolutionary direction of this population. The change of stable equilibrium shows that compared with cooperative AAs, defective AAs are more beneficial to the evolution of cooperation. Other parameters are fixed as $S=0.5$ and $T=1.2$.}
\label{SD}
\end{figure*}

In SD game, when AAs are absent, replicator dynamics have demonstrated that the interior equilibrium $\hat{\rho_C}$ is the unique asymptotically stable state, while the equilibria $\rho_C=0$ and $\rho_C=1$ are always unstable. However, the results will be different if we take AAs into consideration, as indicated in Theorem 2 (electronic supplementary material). We find that the population converges to a full cooperation 
equilibrium state regardless of the initial frequency of human cooperation, provided the condition $\frac{1+y\phi}{1+y} \leq \frac{P-S}{R-T-S+P}$ is satisfied. Interestingly, to achieve a full cooperation equilibrium state with as few AAs as possible, the optimal approach is to introduce AAs with a cooperation probability of $\phi=0$. Correspondingly, the minimum value for $y$ is $\frac{T-R}{S-P}$. On the other hand, when $\frac{y\phi}{1+y} \geq \frac{P-S}{R-T-S+P}$, the population converges to a full defection equilibrium state regardless of the initial frequency of human cooperation. Therefore, we reveal that defective AAs can actually stimulate human cooperation, acting as catalysts for cooperative behavior in SD games.

We then present analytical results regarding how AAs affect the equilibrium by setting $T=1.2$ and $S=0.5$ in Fig.~\ref{SD}.
Panel A depicts the $\phi-y$ phase diagram, which consists of three parts: full cooperation state, full defection state, and coexistence state of C and D. We find that a minority of defective AAs (see the area with $y<1$) can shift the equilibrium state from coexistence state to a full cooperation equilibrium state, whereas a sufficiently large fraction of cooperative AAs drives the system to a full defection equilibrium state (blue color).
To better understand how such an unexpected full defection state happens, we examine the frequency of human cooperation as a function of AAs' cooperation probability $\phi$ at $y=4$ in Fig.~\ref{SD} B. We observe that with a fixed ratio of AAs to human players $y$, the lower the cooperation probability of AAs, the higher the level of human cooperation. In detail, the unique asymptotically stable state moves from full cooperation to the coexistence of C and D, and ultimately to complete defection with the increase of $\phi$.

\subsubsection{Stag hunt game}

\begin{figure*}
\centering
\includegraphics[scale=8]{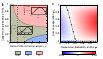}
\caption{ \textbf{Equilibrium and phase diagram in SH game.} (A) Phase diagram showing the stable equilibrium state as a function of the $(\phi, y)$ pair. Introducing cooperative AAs in the population increases the likelihood of achieving a full cooperation equilibrium state, while introducing a large fraction of defective AAs may lead to cooperation collapse. The blue and red areas represent the unique asymptotically stable states of full cooperation and full defection, respectively. The brown area indicates a bistable state that includes both full cooperation and full defection. The insets display phase portraits of $\rho_C$. The dashed line $y=1$ means that human players and AAs each account for 0.5. (B) Stable (solid dots) and unstable (open dots) equilibrium states as a function of $\phi$ with a fixed $y=4$. The background illustrates the gradient of $\rho_C$. Cooperative AAs have a more positive impact on the evolution of cooperation compared to defective AAs. Other parameters are held constant: $S=-0.2$, $T=0.6$.}
\label{SH}
\end{figure*}

In the absence of AAs, replicator dynamics have revealed that the coexistence of C and D is an unstable equilibrium state, whereas the full cooperation and defection equilibrium states are both asymptotically stable. The equilibrium state that the population evolves in depends on the initial frequency of human cooperation.
However, the full cooperation equilibrium state provides each player with a higher payoff compared to the full defection equilibrium state. Consequently, the question arises of how to steer the population towards a full cooperation equilibrium state that is independent of the initial frequency of human cooperation. This can be addressed by incorporating AAs, particularly cooperative AAs (see Theorem 3 in electronic supplementary material). In detail, when $\frac{y\phi}{1+y} \geq \frac{P-S}{R-T-S+P}$, the full cooperation equilibrium state becomes a unique asymptotically stable solution, implying that the population converges to full cooperation irrespective of initial frequency of human cooperation. In particular, to achieve full cooperation with as few AAs as possible, the best option is to introduce AAs with cooperation probability $\phi=1$. On the other hand, the population converges to full defection regardless of the initial frequency of human cooperation when $\frac{1+y\phi}{1+y} \leq \frac{P-S}{R-T-S+P}$. To prevent the collapse of cooperation, it's advisable to control the cooperation probability of AAs with $\phi \geq \frac{P-S}{R-S-T+P}$. 

In Fig.~\ref{SH} A, we present the phase diagram of analytical solutions. There exist three phases, including full C, full D, and a bistable state of C or D. The results demonstrate that the higher the cooperation probability of AAs, the lower the threshold for the proportion of AA required to achieve a full cooperation equilibrium state. Notably, when $y<1$, we showcase that even a minority of cooperative AAs can stimulate a full cooperation equilibrium state. On the other hand, AAs with a lower cooperation probability ($\phi<\frac{P-S}{R-S-T+P}$) can result in the collapse of cooperation in a population containing a large fraction of AAs (see blue area). 
We then show how the equilibrium of the population varies as a function of $\phi$ by fixing $y=4$. In the monostable state, the equilibrium is insensitive to the initial frequency of human cooperation. However, in the bistable state, increasing $\phi$ decreases the unstable interior equilibrium and expands the area capable of reaching a full cooperation equilibrium state.

Overall, we find that a minority of defective (or cooperative) AAs can trigger a pronounced phase transition towards a full cooperation equilibrium state in SD (or SH )games. In contrast, if AAs take a larger proportion, although a full cooperation equilibrium state is easy to reach, there is a risk of transitioning to a full defection equilibrium state. Given the recognition of social learning as a means of describing human strategy updating (\cite{traulsen2010human}), we further investigate the results by considering pairwise comparison rule, focusing on the situation where human players can not only imitate the strategy of human player but also AAs.

\begin{figure*}
\centering
\includegraphics[scale=8.5]{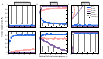}
\caption{\textbf{Equilibrium in three types of social dilemma games via pairwise comparison rule.} In PD and SH games, cooperative AAs benefit cooperation regardless of imitation strength. In SD game, defective AAs promote cooperation effectively under strong imitation strength, aligning with findings in the replicator dynamics. If imitation strength becomes weak, cooperative AAs again stimulate cooperation better. From left to right, each column means the results of PD ($S=-0.2$, $T=1.3$), SD ($S=0.5$, $T=1.2$), and SH ($S=-0.2$, $T=0.6$) games. The lines mean numerical results, and the dot means the agent-based simulation results. The dashed and solid lines represent unstable and stable equilibria, respectively. The upper and bottom panels are obtained with $\phi=0.1$ and $\phi=0.9$, respectively.}
\label{pairw}
\end{figure*}

\subsection{Pairwise comparison rule}

In the pairwise comparison rule, we investigate the influence of AAs as well as the imitation strength $K$. Note that, $K \rightarrow \infty$ and $K \rightarrow 0$ mean strong and weak imitation strength, respectively. The probability given by Fermi function is totally affected by the sign of payoff difference $\pi_C-\pi_D$ if $K \rightarrow \infty$, or tends to 0.5 if $K \rightarrow 0$. 
Since obtaining an analytical solution for pairwise comparison rule is intractable, we present numerical and simulation results in this section. 
A fascinating finding is that results are qualitatively consistent with replicator dynamics if we consider strong imitation strength in pairwise comparison. However, the results vary as the imitation strength weakens. 
We utilize the same values of $T$ and $S$ as in the RD section and present the numerical and simulation results for the pairwise comparison rule in Fig.~\ref{pairw}. Note that we here represent cooperative and defective AAs as $\phi=0.1$ and $\phi=0.9$, respectively. The simulation results are obtained following Algorithm I (electronic supplementary material).

In PD game, human cooperation, either in $\phi=0.1$ or $\phi=0.9$ condition, is difficult to emerge under strong imitation strength (see Fig.~\ref{pairw} A). In detail, when $K=100$, human cooperation is insensitive to the strategy and fraction of AAs, which is consistent with replicator dynamics. However, when we reduce the imitation strength, the results change. Both AAs' proportion and cooperation probability positively influence the evolution of human cooperation. In particular, cooperative AAs promote human cooperation more effectively than defective AAs. This effect becomes more significant with lower imitation strength (see $K=1$). 

In SD game (see Fig.~\ref{pairw} B), human cooperation increases (or decreases) with AA's proportion when $\phi=0.1$ (or $\phi=0.9$) under strong imitation strength. Defective AAs benefit the evolution of human cooperation, which is consistent with the finding in RD. In particular, this effect is insensitive to the initial fraction of human cooperation, as shown in Fig. S1 (electronic supplementary material). 
However, the results become totally contrary when imitation strength weakens (see $K=1$): cooperative AAs are more beneficial to human cooperation.

In SH game (see Fig.~\ref{pairw} C), there exists two kinds of state: a unique asymptotically stable state $\rho_{C1}^*$ (or $\rho_{C2}^*$) and a bi-stable state $\rho_{C1}^*$ and $\rho_{C2}^*$, where $\rho_{C1}^*$ means the coexistence of C and D with a higher frequency of cooperation and $\rho_{C2}^*$ means the coexistence of C and D with a lower frequency of cooperation. As shown in Fig. S2 (electronic supplementary material), in a bistable state, which equilibrium the system evolves to is affected by the initial frequency of human cooperation. Furthermore, similar to the findings in RD, AAs with lower (or higher) cooperation probability $\phi=0.1$ (or $\phi=0.9$) are more feasible to cause the state $\rho_{C2}^*$ (or $\rho_{C1}^*$) as $y$ increases under strong imitation strength. This finding is still robust when imitation strength becomes weak.

\begin{figure*}
\centering
\includegraphics[scale=10]{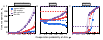}
\caption{\textbf{The frequency of human cooperation as a function of $\phi$ in complex networks with $y=0.95$.} (A) Human cooperation is promoted by AAs in PD game across all network types. In particular, the higher the value of $\phi$, the higher frequency of cooperation. (B) Compared to the results without AAs, human cooperation is prohibited by AAs in SD game regardless of network type. With the increase of $\phi$, human cooperation decreases (or increases) in square lattice (or heterogeneous networks). (C) Cooperative AAs are more beneficial for the evolution of human cooperation. 
The dots represent simulation results averaged over 60 iterations, and the shaded region represents the standard deviation. The dashed lines indicate the frequency of human cooperation in the absence of AA. Parameters are fixed as $S=-0.2$, $T=1.3$ in PD game, $S=0.5$, $T=1.4$ in SD game, and $S=-0.5$, $T=0.6$ in SH game.}
\label{three nets}
\end{figure*}

Our findings demonstrate consistent results in both the replicator dynamics and pairwise comparison rule under strong imitation strength. However, if weak imitation strength (the irrational option of the human player) is taken into account, cooperative AAs are more beneficial to the evolution of human cooperation in all three types of games. While our previous discussions primarily focused on well-mixed populations, exploring the outcomes within networks that incorporate local interactions is also pertinent. 

\subsection{Extension to complex networks}

We first present the average human cooperation frequency $\rho_C$ as a function of $\phi$ for square lattice, Barabasi Albert (BA) scale-free, and Erdos Renyi (ER) random network in Fig.~\ref{three nets}. In PD game, cooperation is consistently promoted with increasing $\phi$, regardless of the network type. The results are qualitatively consistent with previous theoretical analyses. Turning our attention to SD games, we find that AAs inhibit cooperation compared to scenarios without AAs, regardless of the network type. In square lattice, we get a similar conclusion that cooperation is further weakened as $\phi$ increases. However, a contrasting phenomenon emerges when considering heterogeneous networks, such as ER random and BA scale-free networks. We infer that this discrepancy may be attributed to AAs occupying network hub nodes.
To verify it, we conduct simulation experiments on a BA scale-free network under two scenarios (see Fig. S5 in electronic supplementary material):
i) By assigning 4871 nodes (to maintain a similar number of AAs as in Fig.~\ref{three nets}) with the highest degrees as AAs, Fig. S5 presents similar results in Fig.~\ref{three nets} B.
ii) By assigning 4871 nodes with the lowest degrees as AAs, we arrive at the conclusion that defective AAs effectively promote cooperation once again. This finding highlights the significant influence of AAs' location on human cooperation. Next, in SH game, AAs with higher cooperation probability are more beneficial for human cooperation compared to defective AAs. In particular, the increase in $\phi$ triggers tipping points in three types of networks.

\begin{figure*}
\centering
\includegraphics[scale=9]{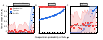}
\caption{\textbf{The frequency of cooperation as a function of $\phi$ in BA scale-free network with ten AAs.} Two scenarios are examined: assigning AAs to the nodes with the largest degree (blue) and assigning AAs to randomly selected nodes (red). Human cooperation closely approximates the results without AAs when AAs are randomly assigned to nodes. However, if AAs are assigned to nodes with the largest degree, even with only 10 nodes, it can stimulate human cooperation effectively, especially in PD and SH games. The dots mean simulation results averaged over 60 times, and the shadow means standard deviation. Parameter settings are the same as Fig.~\ref{three nets}.}
\label{tennodes}
\end{figure*}	

Given the significance of hub nodes, we conducted additional analysis by focusing on ten nodes in three different types of games, as shown in Figure~\ref{tennodes}. When AAs are assigned to nodes with the highest degrees, the strategy of AAs has a profound impact on human cooperation. Even a slight change in the behavior of AAs, especially in PD games (see Figure~\ref{tennodes} A), can lead to a significant increase in overall human cooperation. Conversely, when these ten AAs are randomly allocated across the network, they have little influence on human behavior and exhibit limited utility in altering cooperation levels.

\section{DISCUSSION}

In this study, we investigate human cooperation in hybrid populations, involving interactions between human players and AAs. AAs, who are programmed to choose cooperation with a specific probability, are employed to answer our motivation questions. The human player is assumed to update strategy according to payoff difference given by replicator dynamics and pairwise comparison rule. Theoretic analysis and experimental simulations mainly proceed following well-mixed population and network structures, respectively. Using replicator dynamics, we investigate the impact of AAs on the equilibrium of various social dilemma games, such as the PD, SH, and SD games. We showcase that cooperative AAs effectively promote human cooperation in SH game, but their influence is limited in games with dominant strategies. Surprisingly, in SD game, cooperative AAs can even disrupt cooperation. To achieve a full cooperation state with as few AAs as possible in SH game, introducing AAs with fully cooperating probability is proved to be the most effective approach. Furthermore, our results also show that defective AAs are not useless, as they can stimulate cooperation in SD games. Correspondingly, to achieve a cooperation-dominant state with as few AAs as possible, the best choice is to introduce AAs with always defection. These findings are further verified using pairwise comparison rule with strong imitation strength.
On the other hand, if taking weak imitation strength (which includes irrational options of human players) into account, we demonstrate that cooperative AAs are beneficial for promoting human cooperation regardless of social dilemmas.

In an extended study, we implement experimental simulations involving three types of complex networks. By incorporating spatial structures into the interaction environment, we obtain qualitatively consistent results in the homogeneous network. The differences in heterogeneous networks are mainly due to the location of AAs. By controlling AAs' location, we find that assigning AAs to hub nodes, even in a small proportion, can significantly affect evolutionary outcomes. 
Thus far, we have contributed a model for studying human cooperation in hybrid populations, showing that it is essential to consider environments related to social dilemmas, networks, and imitation strength when designing AAs. The insights gained from our results have practical implications for developing AI algorithms to foster human cooperation.

Our research distinguishes itself from previous studies mainly in several key aspects.
Firstly, in addition to considering the committed minorities (\cite{cardillo2020critical, matsuzawa2016spatial}), we incorporate a substantial number of autonomous agents into our model. These AAs cooperate with their counterparts with a certain probability. The inclusion of a substantial number of AAs is motivated by recent advancements in social network research (\cite{abokhodair2015dissecting}), which suggest that machine accounts make up approximately 32\% of all tweets based on empirical evidence from Twitter data. Moreover, there is an increasing trend in the number of machine accounts, posing significant challenges in terms of reducing their potential risks (\cite{ping2018social}). In the context of human-agent games, we validate that minority cooperative AAs stimulate cooperation, which aligns with existing literature (\cite{xie2011social, centola2018experimental, arendt2015opinions}). However, our research reveals an unexpected result: the inclusion of a large number of AAs leads to a breakdown of the cooperative system, surpassing the effects observed with a mere minority of AAs, as demonstrated in the defective region depicted in the left panel of Fig.\ref{SD} and Fig.\ref{SH}. This discovery emphasizes the potential risks posed by the growing prevalence of AAs in relation to human cooperation.

Secondly, we introduce defective AAs, whose role in fostering cooperation in two-player social dilemma games has been largely overlooked in the context of human-agent interaction. 
Either replicator dynamics or pairwise comparison rule, we find that the inclusion of defective AAs can indeed trigger the dominance of cooperation in SD games, an effect that remains hidden when solely focusing on cooperative AAs. 
Furthermore, although several existing studies primarily use AAs to address fairness or collective risk problems (\cite{terrucha2022art,santos2019evolution}), they have not introduced AAs in structured populations (\cite{nowak1992evolutionary}). These interactive environments have been recognized as important factors in the context of human-human interactions. By introducing structured populations, we reveal tipping points that are triggered by AAs in SH games. Meanwhile, we investigate the effect of nodes with higher degrees on triggering human cooperation. These additional critical extensions provide a comprehensive understanding of the role of cooperative and defective AAs in the evolution of human cooperation. They offer a more realistic representation of interactive environments in human-agent interactions, shedding light on the complex dynamics at play in social systems.

There are still intriguing avenues for future exploration in this field. In human-human interactions, punishment has been proven to be a powerful behavior in eliciting cooperation (\cite{han2016emergence,wang2023emergence}). It's also essential to understand its utility in human-agent populations, particularly in solving second-order free-rider problems (\cite{szolnoki2017second}). Even though theoretical analysis helps identify critical values, it neglects human players' emotional and social factors. Conducting experiments involving structured and unstructured populations to test these findings will open up exciting avenues for research in human-agent interaction.

\subsection{Limitations of the study}

The current study focuses solely on the simplest social dilemmas involving two decisions: cooperation and defection. In social dilemmas where players have no prior information about counterparts, we have shown that AAs with even simple intelligence can facilitate cooperation. However, it still remains uncertain how they would perform in more complex scenarios, such as stochastic games and sequential social dilemma games (\cite{barfuss2020caring, leibo2017multi}). Interaction among players in these scenarios may be influenced by historical information or the state of the environment, making it necessary and meaningful to address such problems. 
Meanwhile, although simple algorithms may be effective in stimulating cooperation (\cite{shirado2020network}), developing algorithms with more intelligence will benefit more to the development of artificial intelligence, especially for human-machine or human-robot interaction (\cite{paiva2018engineering}).
Several subsequent studies have investigated the information level (\emph{i.e.}, to what extent human players know their opponents are agents) on human cooperation (\cite{paiva2018engineering}). These studies revealed that human cooperation fares better in situations where human players have no information about the true nature of their opponents (\cite{ishowo2019behavioural}). Conversely, human players tend to reduce their willingness to cooperate even when participants recognize that agents perform better than human players at inducing cooperation. 
However, Shirado and Christakis conducted the human-agent interactions on network structures and presented contrasting results that human cooperation can still be promoted even if the identity of agents is transparent (\cite{shirado2020network}). In social dilemma games with one-shot settings, there is still no deterministic answer. Our study ignored the intrinsic property of AAs, exploring a scenario without learning bias between humans and AAs. Therefore, it is crucial to consider the true nature of agents, especially in one-shot settings, to expand our understanding of their behavior and implications.

\section{STAR METHODS}	
In this section, we briefly describe the basic concept of social dilemma games and then present our model in the context of hybrid populations.

\subsection{Social Dilemma Games}

Two-player social dilemma game, a typical subclass of social dilemma, depicts the rational decision-making of two participants by introducing a strategy set and payoff matrix. In the simplest version, each player selects a strategy from a strategy set $\mathcal{S}=\{C, D\}$, where $C$ and $D$ represent cooperation and defection, respectively. Mutual cooperation yields a reward $R$ to both players, while mutual defection results in a punishment $P$. Unilateral cooperation leads to a sucker's payoff $S$, while the corresponding defection receives a temptation to defect $T$. The above process can be represented by a payoff matrix: 
\begin{equation}
\mathcal{A} = \left( \begin{matrix}
R & S \\
T & P
\end{matrix} \right).
\label{payoff0}
\end{equation}

Using this payoff matrix, the so-called social dilemma is meeting if it follows these four conditions simultaneously (\cite{macy2002learning}):
	
	i) $R>P$. Players prefer to cooperate with each other than to defect from each other. 
	
	ii) $R>S$. Mutual cooperation is preferred over unilateral cooperation. 
	
	iii) $2R>T+S$. Mutual cooperation is more beneficial for the collective than defecting against a cooperator.
	
	iv) Either $T>R$ (greed) or $P>S$ (fear). The former condition means players prefer exploiting a cooperator to cooperating with him. The latter condition means players prefer mutual defection over being exploited by a defector. 

According to the ranking order of these parameters, these two-player social dilemma games can be classified into four different kinds of games (\cite{wang2015universal}), which are PD games ($T>R>P>S$) where defection is the dominant strategy, SD games ($T>R>S>P$), SH games ($R>T>P>S$), and harmony (H) game ($R>T, S>P$) where cooperation is the dominant strategy. It is noteworthy that the first three games exhibit social dilemmas (\cite{macy2002learning,leibo2017multi}), while the harmony game does not. Without a specific declaration, we set $R=1$ and $P=0$ throughout this paper.


\subsection{Population Setup and Autonomous Agents}
We consider a well-mixed and infinitely large population $\mathcal{P}=\{1, 2, \cdots, N\}$, where $N \rightarrow \infty$, and each player can interact with each other with equal probability. In the population, player $i \in \mathcal{P}$ can choose one of two strategies from set $\mathcal{S}=\{C, D\}$. We denote the strategy of player $i$ as a vector $\mathcal{X}_i = (x_1, x_2)'$, where $x_j=1$ if the $j$th strategy is chosen and the other element is equal to 0. 
To investigate how AAs affect the cooperative behavior among human players, we consider a hybrid population consisting of human players and AAs (see Fig.~\ref{sketch} A). 
Human players actively participate in the game and update their strategies through a social learning process. AAs, on the other hand, follow a pre-designed algorithm to make their choices: they cooperate with a fixed probability $\phi$ $(0 \leq \phi \leq 1)$ and defect otherwise. We refer to them as cooperative AAs if $0.5 \leq \phi \leq 1$, and defective AAs if $0\leq \phi < 0.5$. In the human-agent hybrid population, each player has an equal chance to engage in a two-player social game with other players (see Fig.~\ref{sketch} B). Consequently, the interaction probability between humans and AAs significantly depends on the composition of this population.


\subsection{Hybrid Population Game}

In the hybrid population, the fraction of cooperation among human players is denoted as $\rho_C$ ($0 \leq \rho_C \leq 1$). 
Assuming human players are one unit, add $y$ units of AAs to the hybrid population. Consequently, the fraction of human cooperation is denoted by $f_C = \frac{\rho_C}{1+y}$. 
The parameter $y$ can also be used to quantify the composition of the population: if $0 < y < 1$, it implies that the fraction of AAs is lower than that of human players; whereas $y \geq 1$ indicates a higher proportion of AAs.
Accordingly, the expected payoff of cooperation and defection among human players in a hybrid population can be calculated as follows:
\begin{equation}
\begin{split}
&\pi_C=\frac{1}{1+y}(\rho_C R + (1-\rho_C)S) + \frac{y}{1+y}(\phi R+(1-\phi)S),\\
&\pi_D=\frac{1}{1+y}(\rho_C T + (1-\rho_C)P) + \frac{y}{1+y}(\phi T+(1-\phi)P),
\end{split}
\label{infinite_pay11}
\end{equation}	
where the first term on the right-hand represents the payoff from interacting with human players, and the second term signifies the payoff obtained from interacting with AAs.


\subsection{Replicator Dynamics}

The replicator equation (\cite{roca2009evolutionary}) is a widely used differential equation that depicts evolutionary dynamics in infinitely large populations. Following this rule, the growth of a specific strategy is proportional to the payoff difference. Therefore, the dynamics of human cooperation can be represented by the following differential equation:

\begin{equation}
\begin{aligned}
\dot{\rho}_C & = (1+y) \dot{f}_C \\
& = (1+y) \frac{\rho_C}{1+y} \frac{1-\rho_C}{1+y}(\pi_C-\pi_D) \\
& =\frac{\rho_C(1-\rho_C)}{1+y}(\pi_C-\pi_D),
\label{delta_fc}
\end{aligned}
\end{equation}
where
\begin{equation}
\pi_C-\pi_D = \frac{\rho_C+y\phi}{1+y}(R-T-S+P)+S-P.
\label{eq04}
\end{equation}

By solving $\dot{\rho}_C=0$, we find that there exists two trivial equilibrium $\rho_C=0$ and $\rho_C=1$, and a third equilibrium $\rho_C^*$ that is closely associated with game models and AAs. By solving $\pi_C-\pi_D=0$, one can derive:   
\begin{equation}
\begin{aligned}
\rho_C^* & = \frac{P-S}{P+R-T-S}+y(\frac{P-S}{P+R-T-S}-\phi)\\
& =\hat{\rho}_C+y(\hat{\rho}_C-\phi),
\end{aligned}
\label{eq05}
\end{equation}
where $\hat{\rho}_C=\frac{P-S}{R+P-T-S}$. It is easy to deduce that the restriction $T>R>S>P$ or $R>T>P>S$ guarantees $0<\hat{\rho}<1$. Moreover, $\rho_C^*$ increases with $y$ if $\hat{\rho}_C-\phi>0$, whereas decreases with $y$ if $\hat{\rho}_C-\phi<0$. Since $\rho_C^*$ measures cooperation rate in human players, this equilibrium will vanish if $\rho_C^*<0$ or $\rho_C^*>1$.
Note that $\hat{\rho}_C$ is also the interior equilibrium when the population consists only of human players (\cite{taylor2004evolutionary}), \emph{i.e.}, the scenario $y=0$. Subsequently, the stability of the equilibrium will be discussed from three types of social dilemmas.

We first demonstrate hybrid population dynamics by analyzing replicator equations. As evidence has revealed that human players may update their behaviors through social learning (\cite{traulsen2010human}), one may ask: what results can be obtained when considering pairwise comparison rule? Under this dynamic, we can also examine the effect of imitation strength on the outcomes.


\subsection{Pairwise Comparison Rule}

Pairwise comparison is a well-known social learning mechanism that accurately depicts game dynamics. 
In this process, strategy updating takes place within a randomly chosen pair of players, denoted as $i$ and $j$, with strategy $\mathcal{X}$ and $\mathcal{Y}$ ($\mathcal{X},\mathcal{Y} \in \mathcal{S}$). If $\mathcal{X} \neq \mathcal{Y}$, player $i$ takes $j$ as a reference and imitates its strategy with a probability determined by the Fermi function,
\begin{equation}
W_{\mathcal{X} \leftarrow \mathcal{Y}}=\frac{1}{1+e^{-K(\pi_{\mathcal{Y}}-\pi_{\mathcal{X}})}},
\end{equation}
where $K$ represents selection intensity (which is also known as imitation strength) and measures the irrational degree of human players (or the extent players make decisions by payoff comparisons) (\cite{hauert2005game, sigmund2010social}). In the hybrid population defined, the probability of a cooperator taking a defector as an indicator is given by:
\begin{equation}
\nonumber
	\mathbf{P}_1 = \frac{\rho_C(1-\rho_C+y(1-\phi))}{1+y}.
\end{equation}
Subsequently, the probability that cooperators decrease by one is
\begin{equation}
\begin{aligned}
Q^{-} &= \mathbf{P}_1 W_{C \leftarrow D} \\
&= \frac{\rho_C(1-\rho_C+y(1-\phi))}{1+y}\frac{1}{1+e^{-K(\pi_D-\pi_C)}}.
\end{aligned}
\end{equation}
Similarly, the probability that a defector taking a cooperator as an indicator is 
\begin{equation}
\nonumber
	\mathbf{P}_2 = \frac{(\rho_C+y\phi)(1-\rho_C)}{1+y}.
\end{equation}
Consequently, the probability that cooperators increase by one is
\begin{equation}
\begin{aligned}
Q^{+} &= \mathbf{P}_2 W_{D \leftarrow C} \\
&=\frac{(\rho_C+y\phi)(1-\rho_C)}{1+y} \frac{1}{1+e^{-K(\pi_C-\pi_D)}}.
\end{aligned}
\end{equation}
In total, the dynamics of cooperation can be represented by a master equation 
\begin{equation}
\begin{aligned}
\dot{\rho}_C & = Q^{+} - Q^{-} \\
& = \frac{(\rho_C-\rho_C^2+y\phi-\rho_Cy\phi)e^{K(\pi_C-\pi_D)}}{(1+y)(1+e^{K(\pi_C-\pi_D)})}  - \frac{\rho_C-\rho_C^2+\rho_Cy-\rho_Cy\phi}{(1+y)(1+e^{K(\pi_C-\pi_D)})}.
\end{aligned}
\label{comp_dy}
\end{equation}
We can derive the equilibrium by solving $\dot{\rho}_C=0$.
Since the denominator is larger than 0 evidently, the equilibrium is mainly determined by the numerator. We then denote the numerator as $f(\rho_C)$ and examine the condition: 
\begin{equation}
\begin{split}
& f(0) = y\phi e^{K(\frac{y\phi}{1+y}(R-S-T+P)+S-P)}, \\
& f(1) = y\phi-y.
\end{split}
\label{condition}
\end{equation}
In the presence of AAs, $f(0) \geq 0$ and $f(1) \leq 0$ are established. The equality holds when $\phi=0$ and $\phi=1$, respectively. Therefore, there exists at least one interior equilibrium when $0<\phi<1$. Note that $\phi=1$ is the so-called zealous cooperator (\cite{cardillo2020critical}). Moreover, the imitation strength $K$ also plays a crucial role in the dynamics of human cooperation.

\subsection{Simulation for complex networks}

Building upon the aforementioned results, we further study the network structure effect on human-agent cooperation in this section. Since interaction, in reality, is not limited to well-mixed populations, we also implement experiments in complex networks that contain local interactions. This means that players can only interact with a limited set of neighboring individuals.
To assess the effect of network structure on cooperation, we employ pairwise imitation as a strategy updating rule and measure the expected cooperation rate among human players. 
Following this, players are matched in pairs and imitate their opponent's strategy based on a probability determined by their payoff difference (\cite{ohtsuki2006simple}). We begin by introducing three types of complex networks.

\subsubsection{Network settings}

Denote $\mathcal{G}=\{\mathcal{V}, \mathcal{E} \}$ as a complex network, where $\mathcal{V}=\{1,2,\cdots,N\}$ represents node set, and $\mathcal{E} \subseteq \mathcal{V} \times \mathcal{V}$ is link set. Each node $i\in \mathcal{V}$ represents either a human player or an AA. For the edge $(i, j)\in \mathcal{E}$, each player $i$ is paired up with another player $j$ to play a two-player social dilemma game. We consider a network with $N=10,000$ players and an average degree of $\left \langle k \right \rangle=4$.

\begin{itemize}
\item Square lattice is a homogeneous network. Each player interacts with their four neighbors and receives a payoff by playing with its north, south, east, and west neighbors. It is noteworthy that here we consider lattice with periodic boundary.
\end{itemize}	

\begin{itemize}
\item Barabasi Albert scale-free network is generated following the growth and preferential attach rules (\cite{barabasi1999emergence}). The degree distribution of the ultimate network satisfies a power-law function.
\end{itemize}

\begin{itemize}
\item Erdos Renyi random network is generated by linking two different nodes with a random probability (\cite{erdHos1960evolution}). 
The degree distribution of the ultimate network satisfies the Poisson distribution.
\end{itemize}	


\subsubsection{Agent-based simulation}

We utilized the Monte Carlo simulation to examine the variation of cooperation across different networks. The pseudocode for the simulation is provided in Algorithm II (electronic supplementary material). Initially, each human player is assigned either cooperation or defection with a probability of 0.5, whereas each AA adopts cooperation and defection with probability $\phi$ and $1-\phi$, respectively. With the specific strategy, a randomly chosen player (assumed to be human), denoted as $i$, obtains the payoff by interacting with connected neighbors
\begin{equation}
\label{up}
\mathcal{P}_{i} = \sum_{y \in \Omega_i} \mathcal{X}_i^{T} \mathcal{A} \mathcal{X}_y ,
\end{equation}
where $\Omega_i$ represents neighbor set of player $i$, $\mathcal{A}$ is the payoff matrix given by Fig.~\ref{sketch} B.
After calculating the cumulative payoff, player $i$ decides whether to imitate one of his/her neighbors' strategy with the probability given by the Fermi function
\begin{equation}
W_{\mathcal{X}_i \leftarrow \mathcal{X}_j}=\frac{1}{1+e^{-K(\mathcal{P}_{j}-\mathcal{P}_{i})}},
\end{equation}
where $j$ is a randomly chosen neighbor. We set $K=10$ in the following simulations.
Results are calculated by conducting 60 realizations. For each realization, we fix the total step as 50000, and each value is averaged over 5000 steps when the network reaches an asymptotic state.


 \section*{Acknowledgments}

This research was supported by the National Science Fund for Distinguished Young Scholars (No. 62025602), the National Science Fund for Excellent Young Scholars (No. 62222606), the National Natural Science Foundation of China (Nos. 11931015, U1803263, 81961138010 and 62076238), Fok Ying-Tong Education Foundation, China (No. 171105), Technological Innovation Team of Shaanxi Province (No. 2020TD-013), Fundamental Research Funds for the Central Universities (No. D5000211001), the Tencent Foundation and XPLORER PRIZE, JSPS Postdoctoral Fellowship Program for Foreign Researchers (grant no. P21374). 

\pagebreak

\printbibliography 

\end{document}


	\title{ \LARGE{Electronic Supplementary Material} 
 
 Facilitating Cooperation in Human-Agent Hybrid Populations through Autonomous Agents}
	
	\author{Hao Guo, Chen Shen, Shuyue Hu, et.al.}
	\maketitle
	
  \section{Introduction}

	\setcounter{equation}{0}
	\setcounter{figure}{0}
	\renewcommand\theequation{S\arabic{equation}}
	\renewcommand\thefigure{S\arabic{figure}}

This electric supplementary information provides three theorems of replicator dynamics in the second section. Then, section three reports the algorithm to implement agent-based simulation for pairwise comparison of well-mixed population and complex networks. The last section shows additional results of the model, carried out in order to verify our conclusions.

\section{Theorem}

Through analyzing replicator dynamics, we showcase three theorems:

	\begin{theorem}
		In prisoner's dilemma games ( or harmony games), the equilibrium state $\rho_C=0$ ( or $\rho_C=1$) is always the unique asymptotically stable state regardless of the density of AAs $y$ and their cooperation probability $\phi$.
	\end{theorem}

	\begin{theorem}
		In snowdrift games, in the presence of AAs, i) the equilibrium state $\rho_C = 1$ is the unique asymptotically stable state if $\frac{1+y\phi}{1+y} \leq \frac{P-S}{R-T-S+P}$, ii) the equilibrium state $\rho_C = 0$ is the unique asymptotically stable state if $\frac{y\phi}{1+y} \geq \frac{P-S}{R-T-S+P}$.`
	\end{theorem}


	\begin{theorem}
		In stag hunt games, in the presence of AAs, i) the equilibrium state $\rho_C = 1$ is the unique asymptotically stable state if $\frac{y\phi}{1+y} 
		\geq \frac{P-S}{R-T-S+P}$, ii) the equilibrium state $\rho_C = 0$ is the unique asymptotically stable state if $\frac{1+y\phi}{1+y} \leq \frac{P-S}{R-T-S+P}$.
	\end{theorem}

\section{Agent-based simulation}	

In a well-mixed population, each human player is assigned either cooperation with probability $p$ or defection with a probability $1-p$ initially, where $p$ controls the initial frequency of cooperation. In contrast, each AA adopts cooperation and defection with probability $\phi$ and $1-\phi$ in each round. With the specific strategy, a randomly chosen player (assumed to be human), denoted as $i$, obtains an expected payoff by interacting with other $\mathcal{N}-1$ individuals. Then, $i$ decides whether to imitate the strategy of a randomly chosen individual $j$ who obtains a payoff in the same way according to the Fermi function. The algorithm is given in Algorithm 1. We get the simulation results by setting $N=1000$. On the other hand, the algorithm for simulation in complex networks is given in Algorithm 2.


            
	
	
	

\begin{algorithm}
    \caption{Evolutionary games in a $N$ individual well-mixed hybrid population}
    \label{algor1}
    \renewcommand{\algorithmicrequire}{\textbf{Input:}}
    \renewcommand{\algorithmicensure}{\textbf{Output:}}
    \begin{algorithmic}[1]        
      \REQUIRE a set of $\mathcal{V}=\{1,2,\cdots,N\}$ agent, a symmetric game $G$, the time step $\mathscr{T}$ 
      
    \FOR{each $i \in \mathcal{V}$}
	 \STATE Initialize strategy of human (normal) player $i \in \mathcal{V}$ as C with a fixed probability $p$, and as D with probability $1-p$;\\
     \STATE Initialize strategy of AA $j \in \mathcal{V}$ as C with a fixed probability $\phi$, and as D with probability $1-\phi$;
	\ENDFOR

    \STATE $t\leftarrow 1$\;

     \WHILE{$t<\mathscr{T}$}
        
    \STATE Calculate the expected payoff of strategy $C$ and $D$; \\
    \STATE Choose a human player $i \in \mathcal{V}$ and an arbitrary individual $j \in \mathcal{V}, i \neq j$ randomly;
    \STATE $i$ decides whether to imitate $j$'s strategy via the social learning rule;
    
    $t\leftarrow t+1$\;	
    
    \ENDWHILE
      
    \end{algorithmic}
  \end{algorithm}

	 \clearpage
	


	
	
   
	 	

\begin{algorithm}
    \caption{Evolutionary games of hybrid population in $N$-Agent complex networks}
    \label{algor2}
    \renewcommand{\algorithmicrequire}{\textbf{Input:}}
    \renewcommand{\algorithmicensure}{\textbf{Output:}}
    \begin{algorithmic}[1]        
      \REQUIRE A graph $\mathcal{G}=\{\mathcal{V}, \mathcal{E}\}$ where $\mathcal{V}=\{1,2, \cdots, N\}$ and $\mathcal{E} \subset \mathcal{V} \times \mathcal{V}$, an symmetric game $G$, the time step $\mathscr{T}$  
      
    \FOR{each $i \in \mathcal{V}$}
	 \STATE Initialize strategy of human (normal) player $i \in \mathcal{V}$ with C or D randomly;\\
     \STATE Initialize strategy of AA $j \in \mathcal{V}$ as cooperation with probability $\phi$, and as defection with probability $1-\phi$;
	\ENDFOR

      \WHILE{$t<\mathscr{T}$}
        \STATE $i\leftarrow 1$\;
         \WHILE{$i<\mathscr{N}$}
         
    \STATE Choose individual $j \in \mathcal{V}$ randomly. \\
    
      \IF {$j$ is a human player}          
          
    \FOR{each $k \in \Omega_j$}
    \STATE Individual $j$ and $k$ play the symmetric game and obtain their payoff, respectively. 
    \ENDFOR		
    
    \STATE Choose an agent $m \in \Omega_j$ randomly.

    \STATE Agent $j$ decides whether to learn agent $m$'s strategy via the social learning rule.
    
    \ELSE 
        \STATE $j$ choose cooperation with probability $\phi$, and defection with probability $1-\phi$
    \ENDIF

        \STATE $i\leftarrow i+1$\;
        
        \ENDWHILE
       \STATE $t\leftarrow t+1$\;	
      \ENDWHILE
      
      
      
    \end{algorithmic}
  \end{algorithm}

 
	\section{Results}

\begin{figure*}
		\centering
		\includegraphics[scale=9]{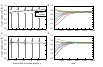}
		\caption{\textbf{The frequency of cooperation as a function of $y$ in pairwise comparison dynamics with strong imitation strength $K=10$.} The cooperation probability of AAs is fixed to $\phi=0.1$ and $\phi=0.9$ in the top and bottom rows, respectively. (a) The frequency of human cooperation $\rho_C$ increases with the increase of AAs' fraction. (c) The frequency of human cooperation $\rho_C$ decreases with the increase of AAs' fraction. (b) and (d) Agent-based simulation with $y=0.1$, the system reaches a unique equilibrium regardless of the initial of $\rho_C$. Parameters are fixed as $T=1.2$, $S=0.5$.}
		\label{tennodes2}
	\end{figure*}

    To support the conclusions in the main text, we show more evolutionary results as follows:

    Fig.~\ref{tennodes2} shows the frequency of cooperation as a function of $y$ in pairwise comparison dynamics with strong imitation strength $K=10$ in the SD game. 

\begin{figure*}
		\centering
		\includegraphics[scale=9]{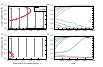}
		\caption{\textbf{The frequency of cooperation as a function of $y$ in pairwise comparison dynamics with strong imitation strength $K=10$.} The cooperation probability of AAs is fixed to $\phi=0.1$ and $\phi=0.9$ in the top and bottom rows, respectively. (a) As the increase of AAs' fraction, the system moves from a bistable state to a monostable state with $\rho_C$ around 0. (b) In bistable region ($y=0.1$), which equilibrium the system reaches depends on the initial of $\rho_C$. (c) As the increase of AAs' fraction, the system moves from a bistable state to a monostable state with $\rho_C$ around 1. (d) In bistable region, which equilibrium the system reaches depends on the initial of $\rho_C$.  As $\phi$ increases, the minimum initial value required to achieve a high cooperation rate gradually becomes smaller. Parameters are fixed as $T=0.6$, $S=-0.2$.}
		\label{tennodes1}
	\end{figure*}

   Fig.~\ref{tennodes1} shows the frequency of cooperation as a function of $y$ in pairwise comparison dynamics with strong imitation strength $K=10$ in the SH game.

\begin{figure*}
		\centering
		\includegraphics[scale=9]{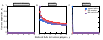}
		\caption{\textbf{The frequency of cooperation as a function of $y$ in complex networks with $\phi=0.1$.} Regardless of the types of the game and network, the frequency of cooperation decreases with $y$. Parameters are fixed as $S=-0.2$, $T=1.3$ in PD game, $S=0.5$, $T=1.4$ in SD game, and $S=-0.5$, $T=0.6$ in SH game.}
		\label{tennodes3}
	\end{figure*}

    Fig.~\ref{tennodes3} shows the frequency of cooperation as a function of $y$ in complex networks with $\phi=0.1$. 
    
\begin{figure*}
		\centering
		\includegraphics[scale=9]{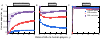}
		\caption{\textbf{The frequency of cooperation as a function of $y$ in complex networks with $\phi=0.9$.} (a) Human cooperation is promoted by AA in the PD game regardless of network types. (b) Human cooperation is prohibited by AA in the SD game regardless of network types. (c) It causes a critical mass effect in ER random and BA scale-free networks with the presence of AA. 
			Parameters are fixed as $S=-0.2$, $T=1.3$ in PD game, $S=0.5$, $T=1.4$ in SD game, and $S=-0.5$, $T=0.6$ in SH game.}
		\label{tennodes}
	\end{figure*}

    Fig.~\ref{tennodes} shows the frequency of cooperation as a function of $y$ in complex networks with $\phi=0.9$. 
    
\begin{figure}
		\centering
		\includegraphics[scale=5.2]{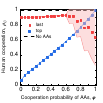}
		\caption{\textbf{The frequency of cooperation as a function of $\phi$ in BA scale-free network, consisting of 4871 AAs.} When AAs are assigned to the nodes with the largest degree, an increase in $\phi$ positively influences human cooperation. Conversely, the trend is reversed if AAs are assigned to nodes with the lowest degree. The dots mean simulation results averaged over 60 times, and the shaded area means standard deviation. Parameters are fixed as $S=0.5$, $T=1.4$.}
		\label{largest network3}
	\end{figure}

    Fig.~\ref{largest network3} shows the frequency of cooperation as a function of $\phi$ in BA scale-free network, consisting of 4871 AAs.

